\documentclass[nofootinbib,floats,aps,superscriptaddress,preprint]{revtex4}
\usepackage{graphicx,bbm,mathdots,feynmf}

\def\lesssim{\mathrel{\mathpalette\vereq<}}

\newcommand{\be}{\begin{equation}}
\newcommand{\ee}{\end{equation}}
\newcommand{\beq}{\begin{eqnarray}}
\newcommand{\eeq}{\end{eqnarray}}

\newenvironment{Eqnarray}
         {\arraycolsep 0.14em\begin{eqnarray}}{\end{eqnarray}}
\def\beqa{\begin{Eqnarray}}
\def\eeqa{\end{Eqnarray}}

\newcommand{\del}{\partial}

\newcommand{\tQCD}{\theta_{\mathrm{QCD}}}
\newcommand{\tQFD}{\theta_{\mathrm{QFD}}}
\newcommand{\twk}{\theta_{\mathrm{weak}}}
\newcommand{\tbar}{\bar{\theta}}

\def\lsim{\mathrel{\rlap{\lower4pt\hbox{\hskip1pt$\sim$}}
    \raise1pt\hbox{$<$}}}         
\def\gsim{\mathrel{\rlap{\lower4pt\hbox{\hskip1pt$\sim$}}
    \raise1pt\hbox{$>$}}}         

\begin{document}

\begin{fmffile}{strongCP20pics}

\preprint{\vbox 
{\hbox{} 
\hbox{} 
\hbox{} 
\hbox{LBNL-56615}
\hbox{UCB-PTH-04/30}
\hbox{hep-ph/0411132}}}

\vspace*{3cm}

\title{\boldmath Strong CP, Flavor, and Twisted Split Fermions}

\vskip 0.3in 

\def\addlbl{Theoretical Physics Group,
Lawrence Berkeley National Laboratory, \\
University of California, Berkeley, CA 94720\vspace*{4pt}}
\def\adducb{Department of Physics, University of California,
Berkeley, CA 94720\vspace*{4pt}}
\def\addlanl{T-8, MS B285, LANL, Los Alamos, NM 87545\vspace*{20pt}}
\author{Roni Harnik}\affiliation{\addlbl}\affiliation{\adducb}
\author{Gilad Perez}\affiliation{\addlbl}
\author{Matthew D. Schwartz}\affiliation{\addlbl}\affiliation{\adducb}
\author{Yuri Shirman}\affiliation{\addlanl}
\vspace*{20pt}

\begin{abstract} \vspace*{8pt}
We present a natural solution to the strong CP problem in the context of split
fermions. By assuming CP is spontaneously broken in the bulk, a weak
CKM phase is created in the standard model due to a twisting in flavor
space of the bulk fermion wavefunctions. 
But the strong CP phase remains zero,
being essentially protected by parity in the bulk and CP on the branes.
As always in models of spontaneous CP breaking, radiative corrections to theta bar 
from the standard model are tiny,
but even higher dimension operators are not that dangerous.
The twisting phenomenon was recently shown to be generic, and not
to interfere with the way that split fermions naturally weaves small numbers
into the standard model. It follows that out approach to strong CP
is compatible with flavor, and we sketch a comprehensive model.
We also look at deconstructed version of this setup
which provides a viable 4D model of spontaneous CP breaking
which is not in the Nelson-Barr class.
\end{abstract}

\maketitle


\section{Introduction}

Among the many mysteries of the standard model, the strong CP problem may
provide a unique window into new physics. The reason is that although the
strong CP phase $\bar{\theta}$ is tiny
($\bar{\theta} \lsim 10^{-10}$
\cite{Peccei:1998jv}),
it is both UV sensitive {\it and} technically natural.
Other tuned UV sensitive parameters,
such as cosmological constant
($\Lambda_{phys} \sim 10^{- 120}$ in units of the Planck mass) correspond to 
relevant 
operators and get large quantum corrections independent of
their initial classical value.  Other technically natural parameters, such as the mass of
the electron
($m_e \sim 10^{- 6}$ in units of the Higgs vev) 
are protected by a custodial symmetry (a chiral symmetry in the case of $m_e$),
and so the quantum corrections are 
proportional to the initial value and thus
calculably small. The strong CP phase has the peculiar property that the
Lagrangian possesses no additional symmetry when $\bar{\theta} = 0$ and yet
the radiative corrections
to $\bar{\theta}$ are still small. In fact, they are so small that from a
model-building point of view,
the strong CP problem is easy: we can just choose by hand a vacuum where $\bar{\theta}=0$ 
at tree level and be done with it. 
But then we are forgoing a promising opportunity. The important point
 is that
while formally $\bar{\theta}$ is UV sensitive, in practice
it acts much like a flavor parameter.
And so by exploring how we may get 
$\bar{\theta}$ to behave exactly like a flavor parameter, we may gain insight into
models of both flavor and CP.

Of course, the real reason to associate the strong CP problem with flavor
physics is much more obvious: all CP violation, weak and strong, can be encoded
in the Yukawa couplings. This is also the reason that the finite
corrections to $\bar{\theta}$ are small: 
CP violation must involve three
generations, and hence many Yukawa couplings
\cite{EllisMaryK,Shab,Khriplovich:1985jr},
most of which are small.\footnote{$\bar{\theta}$ also gets an divergent
logarithmic renormalization at 7-loops, but with a UV cutoff of $\Lambda \sim
M_P$ this turns out to be smaller than the finite piece~\cite{EllisMaryK}.}

So, it is natural to expect a solution to the flavor puzzle, that is a
simplification of the hierarchy of fermion masses, to coordinate with a
solution for the CP problem, that is the hierarchy between
$\theta_{\mathrm{weak}} \sim 1$ and $\bar{\theta}$.  Moreover, precisely
because $\bar{\theta} = 0$ is not protected by a custodial symmetry there
{\it must} be new physics that distinguishes its tree level value from
$\theta_{\mathrm{weak}}$, and this physics may also distinguish the three
generations of fermions.  Yet it is remarkable that most of the solutions to
the strong CP problem involve physics which is not directly related to the
solution of the SM flavor puzzle.

Let us be a little more concrete. 
Although there is only one reparameterization invariant strong CP phase
$\bar{\theta}$, it is useful to write it as:
\begin{equation}
  \bar{\theta} = \tQFD - \tQCD
\end{equation}
Here, $\tQFD$ is related to the phases of the quark masses
\begin{equation}
  \tQFD = \mathrm{Arg\, Det} [ Y_u Y_d ] \label{tqfd}
\end{equation}
and $\tQCD$ is the coefficient of the topological $F \tilde{F}$
term in the QCD Lagrangian:
\begin{equation}
\label{tqcd}
  \mathcal{L} \supset \tQCD \frac{g_s^2}{32 \pi^2} \varepsilon^{\mu \nu
  \alpha \beta} F_{\mu \nu}^a F^a_{\alpha \beta}
\end{equation}
This separation is artificial: a chiral rotation of the quarks will move
the phase between $\tQCD$ and $\tQFD$ through of the chiral anomaly. Thus we can always choose $\tQCD=0$.
But this can be misleading. For example, if we have some flavor model which manipulates the Yukawas
so that $\tQFD=0$, we can no longer assume $\tQCD=0$ by a gauge choice. That is, it is inconsistent to
choose $\tQCD=0$ and {\it then} work on $\tQFD$. We must address the problems together
because the chiral anomaly makes $\tQCD$ and $\tQFD$ indistinguishable.
On the other hand, we can force $\tQCD = \tQFD = 0$.
by imposing a symmetry, such as CP,
instead of simply rotating the phase into $\tQFD$.
Of course, because CP is not a symmetry of the standard
model, it must broken. 
But if it is broken spontaneously, then the corrections to $\tbar$ are finite and calculable,
and the distinction between $\tQCD$ and $\tQFD$ helps us write down operators and estimate their size.
In particular, we can address the strong CP problem through the flavor sector.
The question becomes: after spontaneous symmetry breaking, why is the combination of Yukawas in
$\tQFD$ (\ref{tqfd}) small while the
Jarlskog combination \cite{Jar}
\begin{equation}
\twk= \mathrm{Arg\, Det} [ Y_u Y_d - Y_d Y_u ] \label{tweak}
\end{equation}
is nearly maximal? This has the gist of flavor questions, such as why is the top Yukawa large while the
others are small and hierarchical. So we are led to search for a unified framework to approach the 
strong CP and flavor problems at the same time.

Of the many proposed solutions to the strong CP problem, 
some of which we review 
and compare in Section~\ref{secother}, 
most have no relation to flavor whatsoever. 
For example, the Peccei-Quinn solution~\cite{Peccei:1977hh} 
promotes $\bar{\theta}$ to a field, which dynamically relaxes to zero.
It patently has nothing
to do with flavor.
Another possible
solution is that $m_u=0$, but  this seems to be disfavored by lattice
calculations~\cite{mu0ex}
 (although still conceivable, see e.g. \cite{BNS}).
The models that could relate to flavor all seem to 
involve spontaneous CP breaking, which will be the main subject of this work.

The archetypal spontaneous CP violating theory was designed by
Nelson~\cite{Nelson:1983zb}, and generalized by Barr~\cite{Barr:1984qx}.
It introduces additional scalars and vectorlike fermions which transform non-trivially
under the standard model flavor group. After a scalar get a complex vev which breaks CP,
global symmetries ensure that $\tbar=0$ to leading order. But the symmetries allow $\twk$ to be
large\footnote{A perhaps minimal implementation of Barr's criteria can be found in \cite{Bento:1991ez}.}.
These Nelson-Barr models
bear a strong
superficial resemblance to Froggatt-Nielson~\cite{FN} 
type flavor models.
Indeed,
in the the Froggatt-Nielsen paradigm, new flavor symmetries are also added
to the standard model, along with heavy scalars which break them. The symmetries and breaking
are designed to give the right texture to the standard model CKM.
However, even though Froggatt-Nielsen type
flavor models and the Nelson-Barr type strong CP models have homologous ingredients 
-- they both involve spontaneous symmetry breaking and new heavy scalars --
there does not seem to a compelling model unifying the two.\footnote{
Candidates include \cite{Glashow:2001yz}, which gives the wrong texture and
\cite{Masiero:1998yi}, which involves adjoints of
$SO(3)_{\mathrm{flavor}}$, but has not been developed into an actual flavor model.
\cite{Chang:2004wy}  provides a brief review.}

In the current work, 
instead of trying to merge Nelson-Barr and Froggatt-Nielson,
we mix spontaneous CP breaking with the split fermion approach to flavor physics~\cite{AS}.
The idea behind split fermions is to localize the standard model fields at different places in an extra
dimension, so that overlap of their Gaussian profiles generates hierarchies.
Since the claim-to-fame of the split fermion models is that they do not invoke new symmetries, it
is hard to imagine how they could distinguish the strong and weak CP phases.
However, extra-dimensions do have new symmetries -- 5D Lorentz invariance,
and locality -- which turn out to provide an almost effortless solution of
the strong CP problem. 

\begin{figure}[t]
\centerline{\includegraphics[width=.5\textwidth]{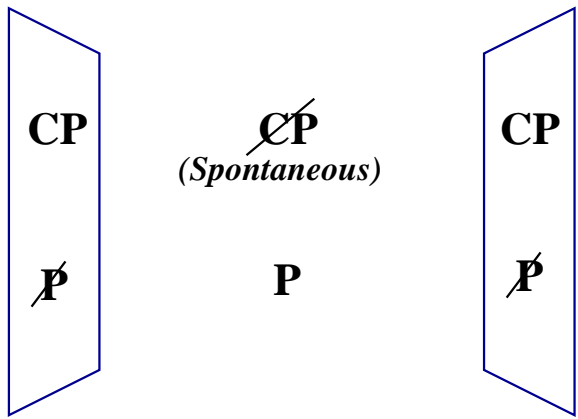}}
\caption{The symmetry breaking structure of our model at leading order.}
\label{figsym}
\end{figure}

Indeed, if we take a 5D orbifold and have CP broken by
a scalar which vanishes on the boundary, where parity is broken,
then either P or CP is a symmetry everywhere, as shown in Figure~\ref{figsym}. 
Because $\tbar$ is P and CP odd, the strong CP phase will vanish at tree level, 
and its radiative corrections are strongly constrained by locality.
In addition, generically the fermion
zero-mode wavefunctions will rotate in flavor space~\cite{GHPSS} as they progress
from brane to brane.
This twist can produce large flavor
mixing and physical CP violating phases, in particular a large
$\twk$, but no $\tbar$. The purpose of paper is to explore
the details of this mechanism.

In Section \ref{sec5D} we review some features of fermions in five dimensions,
including discrete symmetries, splitting and twisting.
Section \ref{5Dmodel} combines these ingredients into a simple model which
shows how twisting can address the strong CP problem. This, without extra
ingredients, sets $\tbar=0$ at leading order and yields $\bar \theta\sim
{\cal O}(10^{-7})$ if we include the strongest higher dimension operators. 
We then show how some non-Abelian flavor symmetries
can be added to reduce $\tbar < 10^{-12}$, which
is well in line with current
bounds.
Some other ways in which the dangerous operators can be suppressed are also discussed.
We then show how the setup is compatible with the split-fermions mechanism for generating
flavor hierarchies.
Section \ref{decon} is devoted to the deconstruction of these
models, where we show the analog of twisting in a purely four-dimensional
context. Although formally the deconstruction works,
the 4D version is much messier than the 5D one, basically
because parity is badly broken. In Section \ref{secother} we show why the use of parity in extra
dimensions is more natural than in left-right symmetric models. 
Finally, in Section \ref{conc} we elaborate our point-of-view, and conclude.


\section{5D Fermions, Splitting and Twisting} \label{sec5D}
The models we will discuss begin in five dimensions on
$M_4\times S_1$
with a set of bulk fermions $\Psi^i$ and scalars $\Phi^a$, 
and perhaps some other fields living on branes.
We will assume CP is an exact symmetry at high energy, and
is spontaneously broken by the condensation of at least one of the bulk scalars
$\langle \Phi\rangle \ne 0$. A scalar with a position dependent vev can be
used to generate fermion hierarchies, and we will see that the same scalar
can be used for both purposes.


In 5D, fermions are vectorlike, and can be written in terms of chiral
fermions as $\Psi = \Psi_L + \Psi_R$, where $\gamma_5 \Psi_L = \Psi_L$ and
$\gamma_5 \Psi_R = - \Psi_R$.  A 5D Lorentz invariant Lagrangian is
automatically invariant under 5D parity, which interchanges $L$ and $R$.  In
particular, a generic Yukawa interaction has the form
\begin{equation} 
\label{lagrangian}
\mathcal{L}_{5D} 
= h_{ij} \Phi \bar{\Psi}_i \Psi_j + \mbox{h.c.}
= \left(h_{ij}\Phi + h_{ij}^{\dagger}\Phi^{*}\right) 
\left(\bar \Psi_{Li}\Psi_{Rj}+ \bar \Psi_{Ri}\Psi_{Lj} \right)\,.
\label{l5d}
\end{equation}
Here, $i$ and $j$ are flavor indices.
After $\Phi$ gets a vev, the effective mass matrix for the fermions is:
\begin{equation}
M_{ij}=  h_{i j}\langle \Phi \rangle +
h_{i j}^{ \dagger} \langle \Phi^{*} \rangle\,.
\label{tilm}
\end{equation}
Note that this matrix is Hermitian, even if the Yukawas $h_{i j}$ are complex.
We emphasize that this is guaranteed by parity which is part of 5D Lorentz invariance.
It is also very different from the way parity works in Left-Right symmetric models,
which we review in Section \ref{secother}.

In order to have a chiral theory, we need to turn our $S_1$ into an orbifold by modding
out by a $Z_2$. The
$Z_2$ acts on fermions as $\Psi(z) \rightarrow
\gamma_5 \Psi(-z)$ and scalars as $\Phi(z) \rightarrow \pm \Phi(-z)$. 
Note that this symmetry forbids a bulk mass term of the form
$m \bar{\Psi} \Psi$, since it sends 
$\bar{\Psi} \rightarrow - \bar{\Psi}\gamma_5$, 
but allows the Yukawa interactions in (\ref{l5d}) 
if the $\Phi$ is  $Z_2$ odd ($\Phi(z) \to - \Phi(-z)$). Such a $\Phi$ must vanish at
the orbifold fixed points and thus its vev will be position dependent~\cite{kink,kink2}.
When we orbifold, by identifying a field with its $Z_2$ image, half the fermion
zero modes, say the right-handed ones, are projected out.

Following from the 5D Lagrangian,
\begin{equation}
  \mathcal{L} = \bar{\Psi}_i 
\left( i \hspace{-.1cm}\not{\hspace{-.07cm}\partial} - 
\gamma^5 \partial_z \right) \Psi_i +
  M_{i j}(z)\bar{\Psi}^i \Psi^j
\end{equation}
the left-handed zero mode profiles are determined by
\begin{equation}
\label{zeromode}
  \left[ \partial_z \delta_{i j} - M_{i j}(z)\right]  \psi_{L\alpha}^j = 0 
\end{equation}
Note that for a generic $z$-dependent matrix $M_{i j}(z)$, 
this is a set of three (for three flavors) coupled differential equations.
The spectator index $\alpha$ distinguishes the three independent solutions and will become the
flavor index in the effective 4D theory.
The form of $M$ and its $z$-dependence, produces both splitting, 
which can establish flavor hierarchies, and twisting, 
which we will use here for strong CP.

Let us write solutions to (\ref{zeromode}) in terms of an evolution operator
$K_{i j}$:
\begin{equation}
\label{sol}
  \psi_L^j ( z ) = K_{i j} (z) \psi^i_L ( 0 )
\end{equation}
where we take $\psi^i_L(0)$ to be an eigenstate of $M_{ij}(0)$. 
It follows that $K_{i j}$ can be written as a Dyson series:
\begin{equation}
  K_{i j} ( z) = P \left\{ e^{\int_0^z M ( z' ) \mathrm{dz}'} \right\} =
  \lim_{N \rightarrow \infty } \prod_n^N \left[ 1 + (z_{n+1} - z_n) M ( z_n ) \right]
\label{pathorder}
\end{equation}
where $P\{\}$ denotes the path-ordered product. An important point is that each
term in this product is an Hermitian matrix, and so
\begin{equation}
\label{det}
\mathrm{Det}[ K_{i j}(z) ] \in
\mathbbm{R}, \qquad \mbox{for each $z$.}
\end{equation}
There is a subtlety here about whether $\psi^i_L$ are orthonormal, and if they are not
whether their diagonalization will introduce a phase into $K$. In Appendix \ref{apportho} 
we show this is not a problem.

In the case that $M_{i j}$ 
can be diagonalized with a $z$-independent flavor rotation, 
the path ordering is trivial. So $K_{i j}$ is diagonalized too. 
Then one can see from the exponential form in 
(\ref{pathorder}) that the zero mode wavefunctions will be localized. Because the
different flavors may be split, that is, localized at different points, their
overlaps can be hierarchical. This is the split fermion solution to the flavor problem.

In general, however, $M_{i j}$ will not be diagonalizable at every $z$ by 
a global flavor rotation. 
In fact~\cite{GHPSS}, if
\begin{equation}
\label{weaker}
\left[ M(z), M'(z)\right] \ne 0, \qquad \mbox{for some $z$}.
\end{equation}
then a zero mode fermion at the boundary $z=\pi$ will differ
from its value at $z=0$ by a twist $K_{ij}(\pi)$.
Note that even if $ M_{i j}(z)$ vanishes at $0$ and $\pi$,
which it will because of the boundary conditions on $\Phi$, the last expression
in (\ref{pathorder}) shows that there will still
be non-trivial twisting because of the intermediate values of $M_{i j}$.
As we will show in the next section, this is exactly what we need
for the CP problem: because $K$ is a complex matrix with real determinant, it can
add a weak CP phase into the Yukawas without adding a strong one.

To be more explicit, CP works in 5D just like in 4D: it acts on fields by Hermitian
conjugation. That is,
$\Phi \rightarrow \Phi^{\star}$ and $\Psi \rightarrow \bar{\Psi}$ 
(and so $\psi_L \rightarrow \bar{\psi}_L$ and $\psi_R \rightarrow \bar{\psi}_R$).
Unlike parity, CP invariance is not an automatic consequence of 5D Lorentz invariance,
but we may impose it. Then, like in 4D, because Lagrangians are already Hermitian, CP
forces all the c-numbers to be real. 
In particular, the Yukawa couplings $h_{i j}$ in (\ref{tilm}) must be real. 

As usual, CP is only defined up to a phase.
So saying CP invariance is equivalent to couplings being real is just a 
convention\footnote{There 
is an interesting model \cite{Masiero:1998yi} of strong CP which uses this ambiguity to show that
one can have ``real CP violation.''}.
Moreover,
we are also free to employ a global flavor rotation to choose our fermion basis, which should
not affect the CP properties of the theory. (This is why it is important to use the basis independent
definitions of $\twk$ and $\tQFD$ in (\ref{tweak}) and (\ref{tqfd}).) 
In the current context, however, there
is an additional subtlety. Even if we cannot use a global flavor rotation to choose an untwisted basis, 
that is one where $M_{ij}(z)$ is diagonal at each $z$, if we can choose a basis so that $K_{ij}$ 
is real, the twist cannot mediate violations of CP. 
The corresponding basis independent condition on $M_{ij}(z)$ is
\begin{equation}
\label{CPbreak2}
\mathrm{Arg\, Det}[M(z),M'(z)] \ne 0, \qquad \mbox{for some $z$.}
\end{equation}
This will be the necessary condition for the twist to mediate CP violation.
Note that this is a stronger condition than that for a twist (\ref{weaker}),
but is still generically satisfied, as we will see in the model in the next section.


\section{A Simple Model for Strong CP \label{5Dmodel}}

We begin with a simple setup which addresses the strong CP problem.
We will find that with the right allocation of fields in the bulk and on the boundary, $\bar \theta$
is naturally zero, but $\twk$ is generically large.
 As always, the finite radiative corrections to $\tbar$ are small. So we go on to consider the
contribution form higher dimensional operators, which cannot be neglected in the non-renormalizable
5D theory. Although we do not give a detailed flavor model, we also show how our flavor hierarchies
can generated within our setup.

\subsection{Tree level solution}
\begin{figure}[t]
\centerline{\includegraphics[width=.45\textwidth]{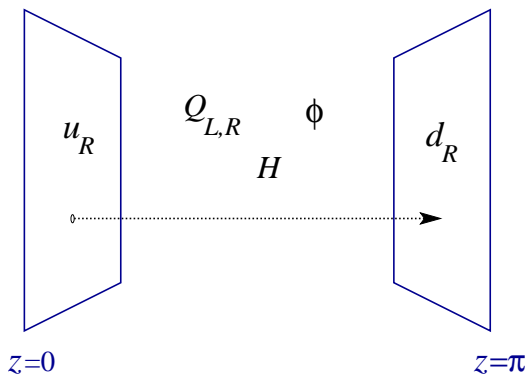}}
\caption{Particle content of the model.}
\label{figsetup}
\end{figure}

As discussed in the previous section, the scene takes place on an orbifold $S^1/Z_2$.
We assume CP is a symmetry at high energy, spontaneously broken by a 
bulk scalar $\Phi$.
The $SU(2)$ quark doublets of the SM, $Q^i$, live in the bulk
while the $SU(2)$ singlet quarks,
$u_R^\alpha$ and $d_R^\alpha$, are 
allocated to opposite fixed points\footnote{The 
main difference between our setup and some others~\cite{GHPSS,GrP} is
that we localize some of the quarks on branes in 5D. This introduces local anomalies, which
can be canceled by adding a parity violating Chern-Simons term in the bulk~\cite{Arkani-Hamed:2001is}.
However, the easiest way to see that local anomalies and bulk parity violation are fictitious
is to think of the singlet quarks as dynamically localized bulk fields.}:
$u_R^\alpha$ at $z=0$ and $d_R^\alpha$ at $z=\pi$.
The right handed component of the bulk
fermions $Q_R^i$ are odd under the orbifold $Z_2$
and thus the right
handed zero modes are projected out of the theory. 
The CP breaking scalar $\Phi$ is a standard model singlet but odd under $Z_2$, so that it
can have Yukawa couplings to the fermions. The SM Higgs $H$ lives in the bulk.
The setup is summarized in Figure~\ref{figsetup}.
%

In addition to the kinetic terms, the Lagrangian of the model, written in chiral 
notation, contains Yukawa  interactions and a scalar potential:
\begin{eqnarray}
\mathcal{L}&\supset&\int d^5x\, \left[(h_{ij}\Phi +h^T_{ij}\Phi^*)
({\bar Q}_{Li} Q_{Rj} + {\bar Q}_{Ri} Q_{Lj})+V_\Phi\right]
\nonumber\\
&+&\int d^4x  \left[ \left.Y^u_{i\alpha} H {\bar Q}_{Li}
u_{R\alpha}\right|_{z=0} 
+  \left. Y^d_{i\alpha} \tilde H {\bar Q}_{Li} d_{R \alpha}\right|_{z=\pi} 
+\mbox{h.c.} \right]
\end{eqnarray}
For clarity, we have used $i$ and $j$ to label the three generations of bulk fields and $\alpha$ to
label fields on the branes. The Yukawa couplings 
$Y^u,\ Y^d$ and $h$ are all real because we assume CP is a good symmetry.

Because the scalar $\Phi$ is odd under the orbifold $Z_2$, its vev must vanish at both
fixed points. However, it may have a non-trivial profile if we include a potential~\cite{GrP}.
Moreover, it is reasonable to give its real and imaginary parts
 different potentials:
\begin{equation}
V_\Phi= \lambda_R[(\Phi+\Phi^*)^2 - \mu_R^2]^2 
+ \lambda_I[(\Phi-\Phi^*)^2 -\mu_I^2]^2
\end{equation}
so that the phase of its vev will generically twist in the extra dimension.

Since the Higgs, $H$, is even under the orbifold $Z_2$,
we will assume its zero mode is flat, for simplicity. Then the
4D Yukawas are determined by the 5D Yukawas and the value of the doublet wavefunctions on the branes:
\begin{equation}
y^u_{\alpha\beta}=Y^u_{\alpha i} Q_{L\beta}^i(0)
\qquad\mbox{and}\qquad
y^d_{\alpha\beta}=Y^d_{\alpha i} Q_{L\beta}^i(\pi)
=Y^d_{\alpha i} K_{ij}(\pi) Q_{L \beta}^j(0), 
\label{yuks}
\end{equation}
Here, $Q_{L \alpha}^i(z)$ are the three vector solutions of equation (\ref{zeromode})
indexed by $\alpha$, with components indexed by $i$.
The evolution matrix $K_{i j}$ is defined in (\ref{pathorder}), and expresses
the effect of twisting on the zero mode profiles.
Because the $Y$'s are real, and because of $K$ has a real determinant, we have
\begin{equation}
\bar{\theta} = \mathrm{Arg Det}( K(\pi) Y^u Y^d ) =0 \label{tbarzero}
\end{equation}
But $K_{i j}$ is complex, because of the complex vev of $\Phi$, and so the twist
induces phases into the Yukawas. 
More precisely, the effective mass matrix in the bulk is
$M=h\langle\Phi\rangle +h^T \langle \Phi^* \rangle$. Since $[h,h^T]\ne 0$ in general, 
and $\langle\Phi\rangle \ne \langle \Phi^* \rangle$,
the criterion (\ref{CPbreak2}) is satisfied and
we should expect $\theta_\mathrm{weak}$, given in (\ref{tweak}),
to be large. So the strong CP problem is solved, at leading order.

\subsection{Corrections}
One might worry that there could be mixing between the zero modes and the higher KK modes of
the fermions, which invalidate our tree level result. 
However, it is not
hard to see that this does not happen. Because of the extra chiral fermions on the brane,
the KK theory has the same number of left-handed and right-handed modes. So we can write down the
effective mass matrices for the up and down type quarks as:
\begin{equation}
\label{KKyukawa}
M^u=
\left( \mbox{
\begin{tabular}{c|ccc}
$Y_{\alpha\beta} \langle H \rangle $ & 0 & 0 & 0 \\ \hline  
 & $\ddots$ & $\vdots$ & $\iddots$ \\
$\vdots$ & $\cdots$ & $M_{KK}^u$ & $\cdots$ \\
 & $\iddots$ &  $\vdots$ & $\ddots$ 
\end{tabular}}
\right),
\quad\quad
M^d=
\left( \mbox{
\begin{tabular}{c|ccc}
$Y_{\alpha\beta} K(\pi)\langle H \rangle $ & 0 & 0 & 0 \\ \hline  
 & $\ddots$ & $\vdots$ & $\iddots$ \\
$\vdots$ & $\cdots$ & $M_{KK}^d$ & $\cdots$ \\
 & $\iddots$ &  $\vdots$ & $\ddots$ 
\end{tabular}}
\right)
\end{equation}
The entries in the top row are the mixings between the left handed KK modes and the right-handed zero modes,
both of which are bulk fields.
The zeros appear because in the KK basis bulk modes at different mass level are orthogonal.
In addition, we cannot write brane-localized mass terms because $Q_R$ vanishes on the brane.
Also, $M^{KK}_u$ and $M^{KK}_d$ are Hermitian, by parity, and thus have a real determinant. It follows
that including the whole KK tower
\begin{equation}
\tbar = \mathrm{Arg\ Det} (M^u M^d) =  \mathrm{Arg\ Det} (K(\pi) Y^u Y^d) =0 .
\end{equation}
So the only relevant quantity for the strong CP problem at tree level are the zero mode Yukawas
appearing in (\ref{yuks}).
Note that the first columns of $M_u$ and $M_d$ are in general non-zero and complex, because the bulk KK modes
do not vanish on the branes, but they do not contribute to the determinant. 


The leading contribution to $\tbar$ in this model comes from higher dimension operators.
Indeed, in 5D model building, one is confronted directly with non-renormalizability
because the gauge interactions blow up near the compactification scale.  For
example, with a 4D gauge coupling $g$,
one cannot take the cutoff higher than
\begin{equation}
\label{cutoff}
\Lambda \sim \frac{24 \pi^3}{L g^2}
\end{equation}
where $L$ is the compactification length. So we must worry
about CP invariant higher-dimension operators which will contribute to $\tbar$
when $\Phi$ gets a vev. 

As shown in Figure~\ref{figsym},
parity forbids a direct contribution to $\tbar$ in the bulk.
But there are dangerous operators which may
couple the bulk to the the branes. For example, consider
\begin{equation}
\label{dangerous1}
\frac{\sqrt{24 \pi^3}}{\Lambda^{7/2}}
i(\del_5 \Phi -\del_5\Phi^*)
F \tilde F\,\delta(z)
\,,
\end{equation}
where the factor of  $\sqrt{24 \pi^3}\Lambda^{-7/2}$ comes from naive
dimensional analysis (NDA) in~5D\cite{NDA}.  This operator gives a direct
contribution to $\tQCD$ once $\Phi$ gets a vev.  If we assume $\langle \Phi
\rangle \sim c L^{-3/2}$ then we get 
\begin{equation} 
\tbar \sim 32 \pi^2
\frac{c\: g^7}{(24 \pi^3)^3} \sim 7 \times 10^{-7} \label{ndaest}
\end{equation} 
where in the last expression we have taken $g \sim 1$ and $c\sim 1$.

We want to emphasize that there is {\it a lot} of unknown physics in this
estimate.  The assumption behind NDA is that the tree level contribution of
higher dimension operators should be comparable to higher loop effect
involving any other operators in the theory.  However, it is possible that
some of the fields are weakly coupled at the cutoff, as is the case for the
photon in the chiral Lagrangian. For example, in our case we may prefer
$\Phi$ to be somewhat weakly coupled.  Recall that $h\langle\Phi\rangle$ is
exponentiated to get the flavor hierarchies, so we should have $h \langle
\Phi \rangle \sim 5 L^{-1}$.  We have taken $\langle\Phi\rangle \sim L^{-3/2}$,
which can be justified by observing that there is already a field with this
vev, namely the radion, so we might as well have two. Then, to get the correct
hierarchy $\Phi$ should couple semi-perturbatively leading to a slightly
smaller estimate for $\tbar$. 

There are other dangerous operators, for example,
\begin{equation}
\label{dangerous2}
\frac{4\pi \sqrt{24 \pi^3}}{\Lambda^{7/2}} \left(
c^1_{\alpha i} \del_5 \Phi + c^2_{\alpha i} \del_5 \Phi^* \right)
H {\bar u}_{R\alpha} Q_{Li}\, \delta(z)
  +\mbox{h.c.}
\end{equation}
We have used the 4D normalization for the 4D field $u_{R \alpha}$, but 5D
normalization for everything else.  This gives contribution to $\tQFD$ and
hence $\tbar$ which is the analog of~(\ref{ndaest}), but smaller by a factor
of $8 \pi$.  One might have expected that~(\ref{dangerous1})
and~(\ref{dangerous2}) should give the same contribution to $\tbar$, because
the two are related by a chiral rotation. However, as discussed in the
introduction and the paragraph before equation~(\ref{CPbreak2}), we have
already fixed a basis in which the Yukawas are real, and so we cannot perform
chiral rotations anymore; the only consistent statement we can make is that
the physical $\tbar$ is given by the larger of the two contributions,
namely~(\ref{ndaest}).

We find it remarkable that without any contrived assumptions, the natural
size of $\tbar$ in this model is small.  We can bring it down to the
experimental constraint by tuning.  In fact, this is what is normally done in
models of spontaneous CP violation.  For example, in 4D models, such as
Nelson-Barr, higher-dimension operators such as (\ref{dangerous1}) are
usually ignored. This is justified, for example in \cite{Nelson:1984hg}, by
taking $\langle \Phi \rangle \ll \Lambda$.  Of course, it is easier to defend
such an approximation in a renormalizable 4D theory than in the 5D case, but
if one includes gravity or any possible UV extension of Nelson-Barr, then the
tuning is just as severe.

If we are not satisfied with fine tuning, there are a number of other
straightforward ways to simply prevent the leading dangerous operators from
appearing at all.  First, note that the dangerous terms involve couplings of
$\Phi$ to brane fields. But we saw in (\ref{pathorder}) that even if the mass
matrix twists far away from the branes, there can still be a large physical
effect. We may thus suppose that $\Phi$ only has support in some region in
the middle of the bulk, and then the CP problem is still solved, but the
dangerous terms identically vanish.  Although we do not know of a way to
justify this, for example through string theory, the important point is that
it turns the strong CP problem over to a completely different type of model
building effort.

Another way around (\ref{dangerous1}) and (\ref{dangerous2}) is to prevent
them with flavor symmetries. For example, suppose instead of a single complex
bulk scalar, we had eight of them, $\Phi_{i j}$ transforming as an adjoint
under $SU(3)_{\mathrm{flavor}}$~\cite{GHPSS}.  This is the diagonal flavor
group of the standard model, under which all the fermions transform as
fundamentals.  The $SU(3)$ invariant Lagrangian has diagonal Yukawas and is
given by: 
\begin{eqnarray}
\mathcal{L}&\supset&\int d^5x\, \Phi_{ij} 
({\bar Q}_{Li} Q_{Rj} + {\bar Q}_{Ri} Q_{Lj})
\nonumber\\
&+&\int d^4x  \left. \delta_{i\alpha} H {\bar Q}_{Li} u_{R\alpha}\right|_{z=0} 
+ \left. \delta_{i\alpha} \tilde H {\bar Q}_{Li} d_{R\alpha}\right|_{z=\pi} 
+\mbox{h.c.} 
\label{adjlag}
\end{eqnarray}
We will assume that the scalar vev rotates in flavor space along the extra
dimension, and so we still have a twist. This assumption may seem implausible
if the scalar potential is flavor symmetric, because the rotation would cost
energy, but we give several examples of how it can work in Appendix
\ref{scalartwist}.

As before, the bulk mass matrix is Hermitian everywhere (now due to the
flavor structure as well as the 5D Lorentz invariance) and the effective 4D
Yukawas again have a real determinant. However, the higher order corrections
are now modified to accommodate the flavor symmetry.  In particular, the
operator (\ref{dangerous1}) is now forbidden.  The leading contribution to
$\tQCD$ now depends on the specific realization of the twist (see Appendix
\ref{scalartwist}), but must be at least second order in $\Phi$. Typically,
we find that the leading contribution to $\tbar$ comes form operators such as
\begin{equation} 
\frac{24 \pi^3}{\Lambda^6}
i {\rm Tr}(\del_5 \Phi)^2 F \tilde F \,
\delta(z).  
\label{dangerous3}
\end{equation} 
which yields
\begin{equation}
\tbar \sim 
32 \pi^2 \frac{c^2 \: g^{12}}{(24 \pi^3)^5}
\sim  10^{-12}
\end{equation}
This is now well out of reach of experimental precision.

Note that with the flavor symmetry, operators like (\ref{dangerous2}), while
not forbidden, no longer contribute to $\tbar$.  Adding such corrections
modifies the Yukawa couplings in~(\ref{adjlag}) from $\delta_{i \alpha}$ to 
\begin{equation}
\lambda^u_{i\alpha}= \delta_{i\alpha} + \left. \frac{ c^1 \del_5
\Phi_{i\alpha} + c^2 \del_5 \Phi^T_{i\alpha}}{\Lambda^{3.5}}\right|_{z=0}
\end{equation} 
and a similar term for the down Yukawas at $z=\pi$.  Even though this is a
complex contribution to the Yukawa couplings, the Yukawa matrix is guaranteed
to be Hermitian by the flavor structure and thus has a real determinant. The
argument for the reality of the determinant of the  effective 4D Yukawa
couplings thus remains intact.

\subsection{Flavor \label{flavor}}
The purpose of this paper is to show that there is a natural solution to the
strong CP problem in the context of split fermions, which is designed to
address flavor. So we will now explain qualitatively how flavor and CP are
naturally incorporated in the same setup. 

In split fermions \cite{AS}, the Yukawas are taken to be order one, and the
flavor hierarchies come from small overlaps of the bulk wavefunctions.  One
might immediately wonder whether twisting would smooth out the hierarchies,
but it was shown in \cite{GHPSS} that this does not happen.  In the current
framework since the $SU(2)$ doublets are in bulk and the singlets on the
brane, it will not be overlap integrals which generate the hierarchies, but
rather the exponential suppression of the zero mode wavefunctions on the
branes.  More precisely, since $u_R$ and $d_R$ are placed on opposite branes,
the size of the effective 4D Yukawa, $y^u$ and $y^d$, will be set by the
value of the $Q_L$ wavefunction at $z=0$ and $z=\pi$ respectively. The
intermediate values of the $Q_L$ wavefunctions are basically irrelevant.

\begin{figure}[t]
\centerline{\includegraphics[width=.4\textwidth]{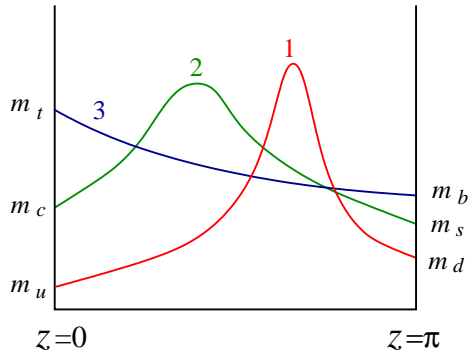}}
\caption{A cartoon showing how mass hierarchies can be generated if only the
SU(2) doublets are in the bulk. $1,2$ and $3$ refer to the norms of the zero
mode wavefunctions for the 3 generations of standard model quarks. }
\label{figflav} 
\end{figure}

The exponential (or Gaussian) behavior of the zero mode wavefunctions can be
seen from the formal solution (\ref{pathorder}). By playing with the order
one factors in the widths and centers of these profiles, we can achieve the
desired structure in the standard model Yukawas. For example, we can arrange
the profile of the first generation to be highly peaked somewhere near the
center of of the extra dimension, so that its value at both branes is small
yielding light $u$ and $d$ quarks. Similarly, the second generation could be
a broader peak, while the third generation is roughly flat. By moving the
centers of the profiles slightly closer to one brane than the other, we can
establish mass differences within a generation, for example, between $u$ and
$d$. See Figure~\ref{figflav}.

Up until now, we have not really shown how the CKM mixing will occur.
Essentially, mixing will be induced due to the twist that introduces a
relative rotation between the mass eigenstates at $z=0$ (up-type) and $z=\pi$
(down type). As mentioned in Section \ref{sec5D}, since this rotation is
generally complex so (\ref{CPbreak2}) is satisfied, there should be an order
one CKM phase. Just to make sure that there is no accidental symmetry which
makes the CKM phase vanish, we have numerically solved the system in a few
simple cases and verified in fact that $\twk$ is large and $\tbar$ is zero.

We furthermore expect that
the hierarchical values of the quark doublet profiles which lead to the mass
hierarchy will dictate the texture of the CKM matrix.  Our framework may thus
be viewed as `predictive' in the sense that the textures of both the mass
hierarchy and the mixing are set only by 6 free parameters. Again, we tried
plugging in a simple guess and numerically evaluated the result. It is not
hard to get a rough quantitative agreement with the standard model; but
this picture, at least qualitatively, gives the correct features.

Finally, we should also mention that the strongest
constraint on flavor models usually comes from flavor-changing neutral currents (FCNCs).
In extra-dimensional models, the strongest constraint comes from the exchange of KK gluons~\cite{FCNC}
which puts a lower bound on the compactification scale: $1/L\lesssim
1000$ TeV. Our estimates for $\tbar$ were basically independent of $L$, and so we have no
problem with simply taking $L$ as small as necessary.


\section{Deconstruction \label{decon}}

In this section we will deconstruct the twisting solution for the CP problem
to obtain a purely 4D mechanism. Our initial motivation was to understand the
twisting within 4D effective theory. However, as we will now show, the 4D model
does not work in the same way as the 5D one. Nevertheless, we end up with a 4D
model of spontaneous CP breaking which is viable but does not fall into the Nelson-Barr 
class.

In practice, deconstructing higher
dimensional theories with fermions is surprisingly tricky.
In the continuum description, one can produce a chiral theory from a 
vectorlike one using 
boundary conditions. On the lattice, an analogous mechanism is not known --
deconstructing bulk fermions directly will always lead to 
the same number of left and right-handed modes.
Thus, in the deconstructed description one needs to {\em start} with a chiral
theory. Moreover, there is also no exact symmetry in 4D corresponding to 5D locality,
so one cannot reliably confine these chiral modes to a particular site.
In practice, one can simulate locality in theory space
with some global symmetries, such as discrete $Z_2$s.
and arrange a mass matrix
that will concentrate the zero modes
on the end cites~\cite{Skiba:2002nx}.
So, in the end, it is not that difficult to create a 4D model which qualitatively matches a
5D one, but we cannot sincerely interpret it as deconstructing the 5D mechanism.




\begin{figure}[t]
\centerline{\includegraphics[width=.3\textwidth]{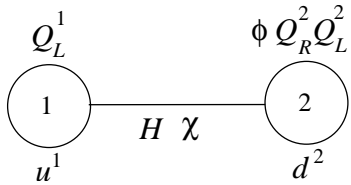}}
\caption{The matter content of our four dimensional model as distributed on
the two sites.}
\label{moose}
\end{figure}

\subsection{Tree level solution}
Let us construct a 2-site model using the ingredients of the 5D solution. 
We start by introducing the matter content of our model, which is summarized
in Figure~\ref{moose}.  The fields are labeled $1$ or $2$ for the first or
second site. As in the 5D model, we want the doublets to live in the bulk and
be vectorlike, so we denote them $Q_L^a$ and $Q_R^a$, $a=1,2$.  To match the
higher dimension model, right-handed bulk fermions should not have zero
modes.  We will achieve this by simply not including $Q_R^1$ into the theory,
thinking of site 1 as a brane.
We will think of site 2 as bulk or brane as necessary.
In the 5D model the singlets, $u$ and $d$,
are on the boundary, so we add only $u^1$ and $d^2$.  The Higgs boson, $H$,
lives in the bulk, thus it has couplings to both sites.  To complete the
matter sector of the model we introduce a complex field $\phi^2$ on the
second site (`in the bulk') and a real link field $\chi$ that connects the
two sites. These fields will play a role similar to $\Phi$, by violating CP
and propagating a twist.

Without enlarging the SM gauge group, or deconstructing it, the
renormalizable couplings between $Q_L^1$ and $d^2$ (or $Q_L^2$ and $u^1$) can
only be prevented through the use of the global symmetries. These symmetries
serve the function of locality in the extra dimension.  A convenient
assignment of $Z_2$ charges is

\begin{center} 
\begin{tabular}{|c|c|c|c|c|c|c|c|c|} \hline
 & $Q_L^1$ & $Q_L^2$ & $Q_R^2$ & $u^1$ & $d^2$ & $\phi$ & $\chi$ & $H$ \\
\hline
$Z_2^1$    & $-$ & $+$ & $+$ & $-$ & $+$ & $+$ & $-$ & $+$\\
\hline
$Z_2^2$    & $+$ & $-$ & $+$ & $+$ & $-$ & $-$ & $+$ & $+$ \\
\hline
$Z^{12}_2$ & $+$ & $+$ & $-$ & $+$ & $+$ & $-$ & $-$ & $+$\\
\hline
\end{tabular} \label{sym}
\end{center}
The most general Lagrangian with this set of discrete symmetries is:
\beq
{\cal L}&=& {\cal L}_{\rm kin}
-\left[
Y_{u}^{ij} \tilde H {\bar Q}^{1}_{Li} u^1_j
+Y_{d}^{ij} H {\bar Q}^{2}_{Li} d^2_j 
+h.c\right] \nonumber\\
&&-\left[\left(\phi h^{ij}+{\phi}^* {h^{ij}_P}\right){\bar Q}_{Li}^{2} 
Q^2_{Rj}+h.c\right]
-\chi f^{ij}{\bar Q}_{Li}^{1} 
Q^2_{Rj} -\chi {f^{ij}}^T {\bar Q}_{Ri}^{2} Q^1_{Lj}
\nonumber\\
&&
-\lambda_R\left[\left({\phi}+{{\phi}^*}\right)^2-\mu^2_{R}\right]^2
-\lambda_I\left[\left({\phi}-{{\phi}^*}\right)^2-\mu^2_{I}\right]^2
-\lambda_\chi\left({\chi}^2-\mu^2_{\chi}\right)^2\,,\label{L2}
\eeq
We also assume CP is symmetry, spontaneously broken 
by the vev of $\phi$.
This forces all the Yukawa couplings to be real.

In addition we choose the Yukawa couplings of $\phi$ to preserve parity on
site 2, that is we take $h_P = h^T$. In contrast to the 5D theory, in which
parity is guaranteed by 5D Lorentz invariance, this parity cannot be taken as
an exact symmetry. It is just a tree-level assumption.  The essential
difficulty is that parity is badly broken in the two site model, where the
bulk and boundaries are indistinguishable.  We will return to this issue in
the next subsection.

Now, once $\phi$ and $\chi$ get vevs (but ignoring the Higgs for now) one
linear combination of the doublets $Q^1$ and $Q^2$ will pick up a mass, and
one will not. The massless combination is determined by the deconstructed
equivalent of (\ref{zeromode}): 
\begin{equation}  
\tilde f^{ij} Q^1_{Lj}+ M^{ij} Q^2_{Lj}=0
\label{dzero} 
\end{equation} 
where
$M^{ij}=\left(\langle\phi\rangle h^{ij}+\langle{\phi}^*\rangle
{h^{ij}_P}\right)$ is the mass at site 2 which plays the role of the hermitian
bulk mass, and $\tilde f^{ij}=f^{ij}\langle\chi\rangle$ is the real link.  In
the case that  $[M,\tilde f]=0$ we can simply choose a flavor basis so that
(\ref{dzero}) is diagonal.  But if  $[M,\tilde f] \ne 0$, the best we can do
is choose a flavor basis so the $Q$'s are diagonal on site one. Then, the
zero modes are \begin{equation} \pmatrix{ \Psi_0 \cr -M^{-1}\tilde f \Psi_0 }
\end{equation} So in this case, $M^{-1}\tilde f$ contains the information
about the non-trivial twist.  In particular, note that the twist matrix
$M^{-1}\tilde f$ has a real determinant, like $K(\pi)$ in (\ref{yuks}),
because $\tilde f$ is real and $M$ is hermitian.  (As in the 5D model, we can
choose an orthonormal basis without introducing a phase into the determinant
of the zero mode, cf. Appendix \ref{apportho}.)

Putting the Higgs back in, we can write the effective Yukawa couplings for the zero modes
as the bare Yukawas times the value of the wavefunctions on the sites:
\begin{equation}
y_u^{\alpha\beta}=Y_u^{\alpha i}\Psi_0^{i\beta} \qquad \mathrm{and} \qquad
y_d^{ab}=Y_d^{\alpha i}(M^{-1}\tilde f \Psi_0)^{i\beta}. 
\end{equation}
This is the 4D analog of (\ref{yuks}), and because the determinant of 
$M^{-1} \tilde f$ is real, $\tbar=0$ at tree level. 
Also, like in the 5D case,
because the Yukawas are in general complex, there will be a generically large
CKM phase.
In fact, the CKM phase is unsuppressed even when the massive modes are
taken to be super heavy.
To see this we can take the limit $|\langle \phi
\rangle| \to \infty$ and $\langle \chi \rangle \to \infty$ such that the
ratio is constant. The twist of the zero mode depends on
$M^{-1}\tilde f$ which remains constant in this limit with an $O(1)$ phase.
So this limit decouples the heavy states and leaves us with just the SM, but with
$\tbar =0$. Note that this is true in 5D too, where $\tbar$ and $\twk$ in the effective
theory are independent of the KK mass scale $L^{-1}$.

We can also see that the massive states do not introduce a contribution to
$\tbar$.  Since the theory is only two sites the full mass matrices for the
up and down-type quarks, the
parallel of~(\ref{KKyukawa}), are only $6\times 6$ and have the form
\begin{equation}
M^u= \pmatrix{ 
v Y_u^{ij} & 0 \cr
{\tilde f}^{ij} & M^{ij}} \qquad
M^d=\pmatrix{ v Y_d^{ij}  & 0 \cr
M^{ij} & \tilde f^{ij}}
\end{equation}
where 
$v$ is the Higgs vev.  The up and down matrices are separated
and are triangular, so we can simply read off that the determinant is real.
Therefore, as in the 5D model, the KK states do not affect that
$\bar\theta=0$ to leading order. 


\subsection{Corrections}
Let us return to the assumption that $h_P=h^T$, which made couplings of
$\phi$ parity symmetric. As mentioned above, this is the analog of 5D Parity,
which made $M$ Hermitian in (\ref{tilm}). However, since parity is not an
exact symmetry, parity violating couplings of $\phi$ will be generated in
perturbation theory. This in turn will generate a strong CP phase. We will
now estimate the minimum natural size of $h_P - h^T$ within the effective
theory.

Due to the exact global symmetries of this  model, the
leading contribution to $h_P-h^T$ arises at two loops. For
example, a correction to the $h \phi {\bar Q}_L^2 Q_R^2$ term
comes from the following diagram:
\begin{equation}
\parbox{50mm}{
\begin{fmfgraph*}(150,100)
\fmfleft{i1,i2}
\fmfright{o1}
\fmf{fermion}{i1,v5}
\fmf{fermion,tension=1/3}{v5,v1}
\fmf{plain}{v1,v2}
\fmf{fermion}{v2,v3}
\fmf{plain}{v3,v4}
\fmf{fermion}{v4,i2}
\fmf{plain,tension=1/4,label=$\phi$}{v4,v5}
\fmf{plain,right,tension=0,label=$\chi$}{v2,v3}
\fmf{plain,tension=1.2}{v1,o1}
\fmfv{l=$Q^2_R$,l.a=120,l.d=.05w}{i1}
\fmfv{l=${\bar Q}_L^2$,l.a=-120,l.d=.05w}{i2}
\fmfv{l=$\phi$,l.a=90,l.d=.05w}{o1}
\fmfv{l=$f$,l.a=-120,l.d=.03w}{v2}
\fmfv{l=$f^T$,l.a=-120,l.d=.02w}{v3}
\fmfv{l=$h^T$,l.a=30,l.d=.03w}{v4}
\fmfv{l=$h$,l.a=-30,l.d=.03w}{v5}
\fmfv{l=$h$,l.a=70,l.d=.03w}{v1}
\end{fmfgraph*} }
\end{equation}
This and other two loop graphs give
\beq 
\tilde h_P(\mu)&=&h^T+ \frac{1}{(16\pi^2)^2}\ln \left( {\mu\over\Lambda}\right)
\left(hf^T f h^T h^T+h^T f^T f h^T h\right)\,\\
\tilde h(\mu)&=&h+ {1\over\left(16\pi^2\right)^2}\ln \left({\mu\over\Lambda}\right)
\left(hf^T f h h^T+h^T f^T f h h\right)\,. \nonumber
\label{L2l}
\eeq
The resulting P violating piece is given by 
\beq
{\tilde h_P(\mu)}^T-\tilde h(\mu)&\propto& h\left[h\,,\,f^T f\right]h^T+ h^T\left[h\,,\,f^T f\right]h
\,.\label{PV}
\eeq
We could have anticipated something like Eq.(\ref{PV}) from a symmetry argument. 
Looking at (\ref{L2}), we see that there is a spurious SU(3) on site 1 under
which $f$ is a fundamental, and so $f f^T$ is an adjoint, and $h$ and $h_P$
are adjoints too.  Thus, ignoring the Yukawas, either $h$ or $ff^T$ alone
would only break this SU(3) down to U(1)$^3$, and so would $h$ and $ff^T$
together if they could be simultaneously diagonalized.  But if there were a
residual U(1)$^3$ symmetry, the system on the first site could be factorized
to three single generation subsystems which could not violate CP and P.  Thus
any contribution to $\tbar$ should contain the commutator factor in
({\ref{PV}}).

Another contribution to parity violation in the couplings of $\phi$ and to
the strong CP phase arises from similar diagrams with $d$ quarks in the loop.
This contribution is the same as above, but has $Y_d Y_d^T$ instead of
$ff^t$.  And so it is a smaller effect, because of the smallness of the down
type Yukawas. Incidentally, this motivates the choice to put the $d$ type
quarks on the second site.

The fact that ({\ref{PV}}) is a two loop effect gives 
$\tbar \sim (16\pi^2)^2 \sim 4 \times 10^{-5}$.
This is quite small, but larger than current bounds.
We can further suppress $\tbar$ by simply tuning
the $\phi$ or $\chi$ Yukawas, $h,f\sim 10^{-2}$.
This is a technically natural tuning because of the residual U(1)$^3$ symmetry
on the first or second site in the limit $f=0$ or $h=0$ respectively.

To summarize, we have shown that the 5D twist has a 4D analog and
can be used to solve the strong CP problem.
The glaring problem with the 4D model 
is that we must tune
$h_P \approx h^T$ 
to achieve the equivalence of 5D parity. 
It is radiatively stable
for these two matrices to differ by one part in $10^{-7}$, but because there is no enhanced
symmetry when $h_P = h^T$, this is not really any better than simply tuning $\tbar$.

It is possible to make the 4D model more technically natural, for example,
by promoting $h_{ij} \phi$ to a field $\phi_{ij}$ transforming in the adjoint of the  flavor group
on site~1.
Then, $\langle \phi_{ij} \rangle$ is guaranteed to be Hermitian without tuning.
Of course, one has to worry about contributions from the explicit flavor violation due to
$Y_d$ and $f$'s, unless these couplings are promoted to fields as well.
One could also construct more elaborate models with more sites
and possibly distribute the gauge group.
This will further suppresses contributions to the CP phase (for
example the leading contribution appears at three loops in three site
model); however, it may also require additional fine-tuning to
guarantee a sufficient amount of twisting.
The bottom line is that we can use this 4D setup to generate a class of solutions
to the strong CP problem which are similar to, but not contained in, the Nelson-Barr
class.
The main drawback of these 4D models is that, in contrast to the 5D models which inspired them,
they do not relate strong CP to flavor.


\section{Comparison with other models \label{secother}}

Our solution to the strong CP problem, based on twisting, 
is related to some other solutions and
it is instructive to explore some of the similarities and differences in more
detail. 
In our 5D model in Section \ref{5Dmodel}, we made essential use of both CP, which we
took to be spontaneously broken, and P, which is a good symmetry in the
bulk and part of 5D Lorentz invariance. The symmetry structure was shown in Figure~\ref{figsym}.
From the 4D point of view, it may seem like we have just produced another Nelson-Barr model,
where the heavy fermions are just KK modes of the bulk fields.
On the other hand, the use of P and CP may remind the reader of left-right (LR) symmetric models~\cite{lrmodels},
which also claimed to have a natural solution to strong CP.
Finally, the twisting looks a little like Hiller-Schmaltz.
In this section, we clarify the distinctions between our model and these others.

\subsection{Nelson-Barr}
Let us start with Nelson-Barr. Nelson's original model was in the context of an $SU(5)$ GUT,
but for our purposes it is simpler to review Barr's generalization.
Call the standard model fermions $F$, and
add some vectorlike generations $R,\bar{R}$ and CP breaking scalars $\Phi$. 
The global symmetries force the mass matrix to look like, in the $F$,$R$,$\bar{R}$ basis,
\begin{equation}
\label{nbmass}
 M_u \sim M_d^T \sim
\left(\begin{array}{ccc}
\mathbbm{R} \langle H \rangle & \mathbbm{C} & 0\\
\mathbbm{C} & 0 & \mathbbm{R} \\
0 & \mathbbm{R} & 0
\end{array}\right) \quad .
\end{equation}
Nelson used a global $U(1)$ to get the top zero and $SO(3)$ flavor to get the
zero in the bottom left, implementing Barr's criteria. The determinant of
this matrix is real, so $\tbar=0$ at tree level, but because of the complex
numbers, $\twk$ will in general be large.

As we have emphasized, radiative corrections to $\tbar$ from the standard
model are small.  If the new fields in Nelson-Barr are to preserve this
situation, there needs to be a separation of scales among the CP-breaking
scalar vev $\langle \Phi \rangle$, the size of the mass term $M {\bar R} R$
for the heavy fermions, and the Higgs vev $\langle H \rangle$. In particular,
we need $\langle \Phi \rangle \gg M \gg \langle H \rangle$. If we ignore the
tuning of the SM Higgs, this reduces to a technically natural tuning of the
fermion masses $M$. So we can see that the virtue of Nelson-Barr is that it
converts the strong CP problem into a technically natural tuning of the
fermion masses $M$. However, there is another tuning implicit in Nelson-Barr:
$\langle \Phi \rangle \ll \Lambda$, where $\Lambda$ is the scale of unknown
new physics. In fact, we should include non-renormalizable operators
suppressed by powers $\Lambda$, such as
\begin{equation}
\frac{(4 \pi \Phi)^n}{\Lambda^n} F{\tilde F}
 \quad\to\quad 
\tbar \sim 32 \pi^2 \left(4 \pi \frac{\langle \Phi \rangle}{\Lambda}\right)^n
\label{nbdanger}
\end{equation}
Here, the $4 \pi$'s have been set by NDA, and the relevant power of $n$
should be determined by the symmetries of the particular theory. But even a
reasonably large $n$, such as  $n=4$,  leads to $\langle \Phi \rangle \lsim
10^{-5} \Lambda$\footnote{It has been pointed out in \cite{nelsonkaplanchoi}
that operators like $\Phi^2 \bar R R/\Lambda$ can also contribute to $\tbar$,
forcing $\langle Phi \rangle < 10^{-10} \Lambda < 10^9$ GeV. Such a low scale
for CP breaking can lead to astrophysically dangerous cosmic strings.}. This
requires a tuning which is less natural than the tuning of $\tbar$, because
the radiative corrections to $\tbar$, unlike $\langle \Phi \rangle$ are
small.  Of course, one can probably get around this tuning by adding new
fields and symmetries, or by including supersymmetry, which allows $\langle
\Phi \rangle$ to be naturally small.

Coming to the comparison with the current work, 
we see that Nelson-Barr matrix~(\ref{nbmass}) looks something like the mass matrix 
we derived in Section~\ref{5Dmodel}. That matrix,~(\ref{KKyukawa}), has the schematic form:
\begin{equation}
M_u= \left( \mbox{
\begin{tabular}{c|c}
$\mathbbm{R} \langle H \rangle$ & 0  \\ \hline  
$\mathbbm{R} \langle H \rangle$ & $\mathbbm{R}$ \\
\end{tabular}}
\right),
\quad\quad
M_d=
\left( \mbox{
\begin{tabular}{c|c}
$\mathbbm{R} \langle H \rangle K $ & 0 \\ \hline  
$\mathbbm{R} \langle H \rangle K $ & $\mathbbm{R}$\\
\end{tabular}}
\right),
\quad\quad
\mathrm{det} (K)\in \mathbbm{R}
\label{ourmm}
\end{equation}
Recall that $K$ is the twist matrix, which is guaranteed to have a real determinant,
ultimately because of 5D Parity. 
In our case, the 0 arises because the heavy fermions are KK excitations of the SM zero modes, and
therefore orthogonal. So not only is (\ref{ourmm}) organizationally different from (\ref{nbmass}),
the underlying symmetries guaranteeing $\tbar=0$ at leading order are not related. Thus, 
although our model involves spontaneous CP violation, it is really not a Nelson-Barr.

In addition, we have shown that non-renormalizable operators which contribute
to $\tbar$ are naturally small. In particular, the operator
(\ref{dangerous3}) gives $\tbar \sim 10^{-12}$ without any tuning at all. In
contrast, the analog (\ref{nbdanger}) with $n=2$ and and say
$\langle\Phi\rangle = 0.01\Lambda$ leads to $\tbar\sim 1$.  The root of the
extra suppression in our case is the symmetry structure shown in
Figure~\ref{figsym}. In particular, since CP is symmetry in the bulk and P is
a symmetry on the brane, locality forces non-renormalizable terms which
violate CP and P (such~as~$F \tilde F$) to involve operators like
$\partial_5$ and $\delta(z)$.  But these turn into factors of $(\Lambda
L)^{-1} \ll 1$ in the low energy theory.

\subsection{Left-Right models}
In addition to Nelson-Barr, our model has much in common with, but is different from, 
left-right symmetric models\cite{lrmodels}.
Left-right models are based on the symmetry group $\mathrm{SU}(2)_L \times \mathrm{SU}(2)_R$.
The fermions $Q_L^i$ and $Q_R^i$ are fundamentals and the Higgs is a 
bifundamental $\Phi$.
Under parity $Q_L^i \leftrightarrow Q_R^i$ and $\Phi \leftrightarrow \Phi^{\dagger}$ 
($\Phi$ should be thought of as a $2\times 2$ matrix, and $i$ is a flavor index).
If we take a generic Yukawa interaction $f_{ij} \Phi {\bar Q}_i Q_j$ and add its
parity and Hermitian conjugate, we get
\begin{equation}
  L \supset ( f_{i j} + f^{\dagger}_{i j} ) \Phi \bar{Q}_L^i Q_R^j + ( f_{i j}
  + f^{\dagger}_{i j} ) \Phi^{\dagger} \bar{Q}_R^i Q_L^j
\label{lrlag}
\end{equation}
One actually needs separate Yukawas for $\Phi$ and 
$\sigma_2 \Phi^* \sigma_2$ to have a realistic flavor model, and additional fields to break parity, 
but these additions are irrelevant here.
To solve the strong CP problem,
$\Phi$ needs to get a vev so that the effective mass matrix $M = ( f +
f^{\dagger} ) \langle \Phi \rangle $ has a real determinant. 
There are two possibilities \cite{Mohapatra:1997su}: 
1) If $f$ is real,
by CP, but $\langle \Phi \rangle $ is complex, breaking CP
spontaneously, then $\det M$ will be complex, so that does not work.
2) Instead, we can take $f$ complex but $\langle \Phi \rangle$ real, which does
lead to an Hermitian $M$; but if $f$ is complex, CP is not a symmetry and then there is no reason
to expect $\langle \Phi \rangle $ to be real. 
This is not to say that left-right models cannot solve strong CP
(for example, there are natural left-right solutions which invoke supersymmetry\cite{Mohapatra:1996xd}),
only that $\tbar$ is not zero automatically. 

In contrast, recall that our
Yukawa interactions are of the form (\ref{lagrangian})
\begin{equation}
  L \supset ( f_{i j} \Phi + f^{\dagger}_{i j} \Phi^{\star} ) \bar{Q}_L^i
  Q_R^j + ( f_{i j} \Phi + f^{\dagger}_{i j} \Phi^{\star} ) \bar{Q}_R^i Q_L^j
\label{ourcoups}
\end{equation}
So if $f$ is real, but $\langle \Phi \rangle $ complex, which is natural,
then the effective mass matrix will automatically be Hermitian, and so the determinant is
real. The difference is that our $\Phi$ is a singlet and not a bifundamental
and so the couplings in (\ref{ourcoups}) are allowed.


\subsection{Hiller-Schmaltz}
Finally, it is insightful to observe a formal similarity between our model
and that of Hiller and Schmaltz \cite{Holdom:1999ny,HS}. 
Hiller-Schmaltz is a supersymmetric model, in which CP is spontaneously broken at a much
higher scale than SUSY. Due to non-renormalization theorems in SUSY, $\tQCD$ does not get renormalized.
All the CP violation in the low energy theory enters through finite contributions to
wavefunction renormalization factors $Z_i$. Because of the reality of the Kahler
potential these factors come through Hermitian matrices with real
determinants. Thus they contribute to a CKM phase but not to $\tQFD$, and so $\bar\theta$
remains zero. 

To see how this relates to our model, recall that in our case, the CP violation comes in through $z$-dependence
of the bulk fermion masses.  If we preform a $z$-dependent flavor rotation
to diagonalize these masses, undoing the twist, the CP violation appears in $z$-dependent kinetic terms.
As in Hiller-Schmaltz, the coefficient of the kinetic terms are Hermitian,
and so the kinetic terms generate a CKM phase, but no $\bar\theta$. Actually, the
analogy is not quite so clean, because the coefficient of the kinetic terms in the low energy theory
will involve the twist matrix $K$ (\ref{det}), which is not Hermitian, but still has a real determinant.
And of course, in our model the phase rotation contributing to CP is a classical effect, 
while in Hiller-Schmaltz it comes from loops. But still, it is entertaining to consider
the similarity between the two models, especially in light of AdS/CFT, which also relates 4D RG flows to
moving along an extra dimension.
Finally, we should comment that both Hiller-Schmaltz and our model solve the strong CP problem
within existing frameworks, SUSY and split fermions respectively, which have been established for other purposes.


\section{Discussion \label{conc}}
Whether the strong CP problem is a serious inadequacy of the standard model
depends on your attitude towards fine tuning. On the one hand, setting
$\tbar=0$ is not natural in the 't Hooft sense because no symmetry is
enhanced in that limit. On the other hand, $\tbar$ is stable at $10^{-14}$
and so it is technically natural in the ``effective field theory sense;''
that is, if one imposes a cutoff (say $\Lambda \sim M_{\mathrm{Pl}}$),
the radiative corrections are smaller than its initial value, 
even without any additional symmetries.
If we believe that the second type of tuning is acceptable, there is no strong CP problem,
because we can just set $\tbar=0$ at the cutoff. However, if we allow only the
first type of tuning, we should attempt to make $\tbar=0$ 't Hooft natural
by imposing a symmetry, such as CP. Then the strong CP problem becomes: why is
weak CP violation in the standard model so large?

Realistic models of spontaneous CP violation which answer this  question
often invoke many new particles and new
global symmetries, and eventually distribute the tuning of $\tbar$
among tunings of Yukawa couplings. But if one accepts that there should
be new physics at some cutoff, then 
as we showed in Section~\ref{secother}, one
must also confront the tuning of scalar vevs.
Such tunings are UV-sensitive and not technically natural in any sense,
and therefore worse than simply setting $\tbar$ to zero by hand. Although
one should be able to avoid the tunings by adding more structure to the model,
the point remains that most models in this class are designed
{\it just} to solve the strong CP problem.

In this paper, we not only present a solution to the strong CP problem 
with minimal tuning, we do so in the context of flavor. We have shown that with a simple setup
-- a 5D orbifold with the SU(2) doublets in the bulk, the singlets on
opposite branes, and a single CP breaking scalar and Higgs -- 
$\tbar$ comes out at $10^{-7}$. No additional symmetries are assumed.
This estimate is based on the most
general assumptions about the UV completion; the contribution from the standard model
is much smaller.
Moreover, with parameters of order one that shift
the bulk wavefunctions, exponentially varied masses and mixing angles
can be generated using split-fermions.
Of course, $\tbar$ currently looks to be smaller
than $10^{-10}$, and so either some tuning, or a little more model
building is necessary. We have suggested some ways in which this may be done.

We have also produced a 4D realization of the twisting mechanism for breaking
CP. However, in order to reproduce the effect of exact parity in the bulk one
must either start with a tuned model at tree level, or assume some flavor
symmetries.  The simplest such realization is a two site model that gives an
acceptable strong CP phase once some Yukawa couplings are assumed to be small
$\sim 10^{-2}$. This model bears resemblance to conventional Nelson-Barr type
models, however we have shown that it is not actually in the Nelson-Barr class.

There is much more to be done. We would like to know if a flavor model based
on this setup is compatible with CKM parameters without tuning.  To make
progress, we would need to know more about how to choose the potential for
the bulk fields so that they produce acceptable fermion profiles and
sufficient twisting.  If we assume flavor symmetries, we would need to know
how the flavor symmetries are broken to produce a twist.  Also, it would be
interesting to look at the cosmology of these models, in particular at
baryogenesis. It was noted in~\cite{NaPe} that this framework may satisfy some of
the Sakharov conditions-- it contains a scalar that may undergo a first order
phase transition, and it has a new source of CP violation. 
So twisted split fermions
potentially addresses all of
the CP issues in the standard model, strong, weak, and baryogenesis, as
well as the flavor puzzle.

\begin{acknowledgments}
The authors are especially grateful to Raman Sundrum for many crucial insights, and
to Yasunori Nomura and Eduardo Ponton for helpful discussions. 
The research of Y.S. is supported
by is supported by the Department of Energy, under contract W-7405-ENG-36 and
Feynman Fellowship at LANL. GP, MS and RH are supported in part by the DOE
under contract DE-AC03-76SF00098 and in part by NSF grant PHY-0098840.
\end{acknowledgments}

\appendix
\section{Orthonormalization \label{apportho}}
In this appendix we will address the following subtlety. 
The zero mode wavefunctions $\psi^i$, the solutions to (\ref{zeromode}),
are orthogonal only when we integrate over the extra dimension
\begin{equation}
\int_0^\pi \psi^i_\alpha(z)\psi^j_\alpha(z)\, dz = \delta^{i j}
\end{equation}
Recall that $\alpha$ labels which state and $i$ its direction in flavor space.
But the solutions we used, given by (\ref{sol}) and (\ref{pathorder}), do not necessarily satisfy
this requirement. The concern is that by taking orthonormal linear combinations, by a transformation
such as
\begin{equation}
\psi_\alpha^i(z)_ \to \psi_\beta^i(z) O^\beta_\alpha
\end{equation}
we might introduce a strong CP phase. We will now show that this does not happen.

The orthonormalization procedure can be done in two steps:
we first multiply by a real diagonal matrix, $O_1$, such that all three
solutions $\psi^i$ are normalized to 1 when integrated over the extra
dimension. We then multiply by a second matrix $O_2$ that will render the
solutions orthogonal. Without loss of generality this matrix may take the
form
\beq
O_2=\pmatrix{1&\sin t e^{is} &0\cr0& \cos t&0\cr 0&0&c}\cdot
\pmatrix{1&0 &\sin w e^{i u}\cr0&1&\cos w\sin x e^{i v}\cr0  &0&\cos w \cos x
}
\eeq
where the first matrix in the product makes $\psi^2$ orthogonal to $\psi^1$,
and the second makes $\psi^3$ orthogonal to $\psi^1$ and the new $\psi^2$. 
Because both $O_1$ and $O_2$ have real determinants,
the orthonormalization procedure does not introduce a
phase into the determinant of the profile matrix, and thus does not damage our
solution to the strong CP problem.

\section{Twisting with Symmetries}
\label{scalartwist}

In section \ref{flavor} we mentioned that it may
be useful for the bulk scalar to be an adjoint of the flavor group.
We then assumed that the vev of the bulk scalar could be twisted.
However, since we impose a flavor symmetry, the bulk potential for the scalar
must be flavor invariant. What causes the adjoint to pick a different
direction in flavor space in different locations? One would expect that a
single adjoint would tend to be aligned, 
but also that twisting along the flat direction  costs little energy.
In this appendix we show some examples where a twist generically occurs. 

One possibility is to add more adjoints with a
non-trivial potentials such that they tend to increase at different rates.
For example, with two adjoints $\Phi_{1,2}$, if the potential contains a term such as 
\begin{equation}
\label{V12}
V_{12}=\left[\mathrm{Tr}(\Phi^1\Phi^2)\right]^2,
\end{equation}
then the adjoints
will prefer to point in perpendicular directions in SU(3) flavor space. 
If they point in commuting directions, of which there are two
in SU(3),
no twisting will be induced. But if we add a third adjoint with similar couplings
then it cannot choose an additional orthogonal direction
which is diagonal.

Now, the bulk fermions will couple to an arbitrary combination of
these adjoints such that the effective mass is
\begin{equation}
M_{ij}(z)=\sum_A c_A \Phi_{ij}^A(z).
\end{equation}
If the adjoints indeed increase at different rates and do not commute, then
this mass matrix is twisted. In general, if the number of adjoints is larger than
the rank of the group, a twist will be induced. It is likely that fewer adjoints are
necessary, but we merely wanted to present sufficient conditions that it could be done.

A second possibility is to force a twist by violating the flavor symmetry
explicitly on both boundaries.
Suppose that each boundary breaks the flavor group down to an
SU(2)
subgroup.
This breaking can be
communicated to the bulk by introducing a scalar field with even orbifold parity
and flavor indices, say an adjoint $\chi$.
Since it is even it cannot couple to the fermions in the bulk, and thus
hardly affects the zero modes. It can, however, couple to the CP breaking adjoint $\Phi$,
for example through a term like $Tr[\chi\Phi\Phi]$.
An arbitrary potential can be now written for $\chi$  on the
boundaries. This will induce twisting of its vev, which will in turn twist $\Phi$.


\end{fmffile}

\begin{thebibliography}{01}
\vspace*{3mm}

\bibitem{Peccei:1998jv}
see e.g. R.~D.~Peccei,
arXiv:hep-ph/9807516;
Y.~Nir,
arXiv:hep-ph/0109090.
P.~G.~Harris {\it et al.},
Phys.\ Rev.\ Lett.\  {\bf 82}, 904 (1999).

\bibitem{EllisMaryK}
J.~R.~Ellis and M.~K.~Gaillard,
Nucl.\ Phys.\ B {\bf 150}, 141 (1979).

\bibitem{Shab}
E.~P.~Shabalin,
Sov.\ J.\ Nucl.\ Phys.\  {\bf 28}, 75 (1978)
[Yad.\ Fiz.\  {\bf 28}, 151 (1978)].

\bibitem{Khriplovich:1985jr}
I.~B.~Khriplovich,
Phys.\ Lett.\ B {\bf 173}, 193 (1986)
[Sov.\ J.\ Nucl.\ Phys.\  {\bf 44}, 659.1986\ YAFIA,44,1019 (1986\ YAFIA,44,1019-1028.1986)].

\bibitem{Jar}
C.~Jarlskog,
Phys.\ Rev.\ Lett.\  {\bf 55}, 1039 (1985).

\bibitem{Peccei:1977hh}
R.~D.~Peccei and H.~R.~Quinn,
Phys.\ Rev.\ Lett.\  {\bf 38}, 1440 (1977).


\bibitem{mu0ex}
S.~A.~Gottlieb,
Nucl.\ Phys.\ Proc.\ Suppl.\  {\bf 128}, 72 (2004)
[Nucl.\ Phys.\ Proc.\ Suppl.\  {\bf 129}, 17 (2004)]
[arXiv:hep-lat/0310041];
A.~C.~Irving, C.~McNeile, C.~Michael, K.~J.~Sharkey and H.~Wittig  [UKQCD
                  Collaboration],
Phys.\ Lett.\ B {\bf 518}, 243 (2001)
[arXiv:hep-lat/0107023].



\bibitem{BNS}
T.~Banks, Y.~Nir and N.~Seiberg,
arXiv:hep-ph/9403203.


\bibitem{Nelson:1983zb}
A.~E.~Nelson,
Phys.\ Lett.\ B {\bf 136}, 387 (1984).

\bibitem{Barr:1984qx}
S.~M.~Barr,
Phys.\ Rev.\ Lett.\  {\bf 53}, 329 (1984).

\bibitem{Nelson:1984hg}
A.~E.~Nelson,
Phys.\ Lett.\ B {\bf 143}, 165 (1984).

\bibitem{Bento:1991ez}
L.~Bento, G.~C.~Branco and P.~A.~Parada,
Phys.\ Lett.\ B {\bf 267}, 95 (1991).


\bibitem{FN}
C.~D.~Froggatt and H.~B.~Nielsen,
Nucl.\ Phys.\ B {\bf 147}, 277 (1979).

\bibitem{Glashow:2001yz}
S.~L.~Glashow,
arXiv:hep-ph/0110178.


\bibitem{Masiero:1998yi}
A.~Masiero and T.~Yanagida,
arXiv:hep-ph/9812225.

\bibitem{Chang:2004wy}
D.~Chang and W.~Y.~Keung,
arXiv:hep-ph/0403070.

\bibitem{AS}
N.~Arkani-Hamed and M.~Schmaltz,
Phys.\ Rev.\ D {\bf 61}, 033005 (2000)
[arXiv:hep-ph/9903417].

\bibitem{GHPSS}
Y.~Grossman, R.~Harnik, G.~Perez, M.~D.~Schwartz and Z.~Surujon,
arXiv:hep-ph/0407260.

\bibitem{kink}
H.~Georgi, A.~K.~Grant and G.~Hailu,
Phys.\ Rev.\ D {\bf 63}, 064027 (2001)
[arXiv:hep-ph/0007350];
\bibitem{kink2}
G.~Perez,
Phys.\ Rev.\ D {\bf 67}, 013009 (2003)
[arXiv:hep-ph/0208102].

\bibitem{GrP}
Y.~Grossman and G.~Perez,
Phys.\ Rev.\ D {\bf 67}, 015011 (2003)
[arXiv:hep-ph/0210053].

\bibitem{Arkani-Hamed:2001is}
N.~Arkani-Hamed, A.~G.~Cohen and H.~Georgi,
Phys.\ Lett.\ B {\bf 516}, 395 (2001)
[arXiv:hep-th/0103135].

\bibitem{NDA}
Z.~Chacko, M.~A.~Luty and E.~Ponton,
JHEP {\bf 0007}, 036 (2000)
[arXiv:hep-ph/9909248].


\bibitem{FCNC}
A.~Delgado, A.~Pomarol and M.~Quiros,
JHEP {\bf 0001}, 030 (2000)
[arXiv:hep-ph/9911252];
D.~E.~Kaplan and T.~M.~Tait,
JHEP {\bf 0111}, 051 (2001) [arXiv:hep-ph/0110126];
G.~Barenboim,  {\it et. al.},
Phys.\ Rev.\ D {\bf 64}, 073005 (2001)
[arXiv:hep-ph/0104312];
W.~F.~Chang, I.~L.~Ho and J.~N.~Ng,
Phys.\ Rev.\ D {\bf 66}, 076004 (2002)
[arXiv:hep-ph/0203212].

\bibitem{Skiba:2002nx}
W.~Skiba and D.~Smith,
Phys.\ Rev.\ D {\bf 65}, 095002 (2002)
[arXiv:hep-ph/0201056].



\bibitem{lrmodels}
J.~C.~Pati and A.~Salam,
Phys.\ Rev.\ D {\bf 10}, 275 (1974);
R.~N.~Mohapatra and J.~C.~Pati,
Phys.\ Rev.\ D {\bf 11}, 566 (1975);
G.~Senjanovic and R.~N.~Mohapatra,
Phys.\ Rev.\ D {\bf 12}, 1502 (1975).

\bibitem{nelsonkaplanchoi}
K.~w.~Choi, D.~B.~Kaplan and A.~E.~Nelson,
Nucl.\ Phys.\ B {\bf 391}, 515 (1993)
[arXiv:hep-ph/9205202].


\bibitem{Mohapatra:1997su}
R.~N.~Mohapatra, A.~Rasin and G.~Senjanovic,
Phys.\ Rev.\ Lett.\  {\bf 79}, 4744 (1997)
[arXiv:hep-ph/9707281].

\bibitem{NaPe}
Y.~Nagatani and G.~Perez,
arXiv:hep-ph/0401070.

\bibitem{Mohapatra:1996xd}
R.~N.~Mohapatra and A.~Rasin,
Phys.\ Rev.\ Lett.\  {\bf 76}, 3490 (1996)
[hep-ph/9511391];
R.~Kuchimanchi,
Phys.\ Rev.\ Lett.\  {\bf 76}, 3486 (1996)
[arXiv:hep-ph/9511376].


\bibitem{Holdom:1999ny}
B.~Holdom,
Phys.\ Rev.\ D {\bf 61}, 011702 (2000)
[arXiv:hep-ph/9907361].

\bibitem{HS}
G.~Hiller and M.~Schmaltz,
Phys.\ Rev.\ D {\bf 65}, 096009 (2002)
[arXiv:hep-ph/0201251];
G.~Hiller and M.~Schmaltz,
Phys.\ Lett.\ B {\bf 514}, 263 (2001)
[arXiv:hep-ph/0105254].




\end{thebibliography}
\end{document}